\documentclass[journal,a4paper]{IEEEtran} % use larger type; default would be 10pt
\usepackage{caption}
\usepackage{graphicx} % support the \includegraphics command and options
\graphicspath{{figures/}} %Setting the graphicspath

\usepackage{subcaption} % make it possible to include more than one captioned figure/table in a single float
\usepackage{amsmath}
\usepackage{cite}
\usepackage{float}

\title{Effects of Turbulence Induced Scattering on Underwater Optical Wireless Communications}
\author{\IEEEauthorblockN{Callum T. Geldard,
		John Thompson,~\IEEEmembership{Fellow,~IEEE}~and~Wasiu O. Popoola,~\IEEEmembership{Senior Member,~IEEE}}% <-this % stops a space
	\vspace{0.1cm}%
\thanks{The authors are with the School of Engineering, Institute for Digital Communications, The University of Edinburgh, Edinburgh EH9 3JL, U.K. (e-mail: callum.geldard@sms.ed.ac.uk; j.s.thompson@ed.ac.uk; w.popoola@ed.ac.uk).
}}
\begin{document}
\maketitle
\begin{abstract}
This paper presents a comprehensive description of the relative effect of optical underwater turbulence in combination with absorption and scattering. Turbulence induced scattering is shown to cause and increase both spatial and temporal spreading at the receiver plane. It is also demonstrated that the relative impact of turbulence on a received signal is lower in a highly scattering channel. Received intensity distributions are presented confirming that fluctuations in received power from this method follow the commonly used Log-Normal fading model. The impact of turbulence induced scattering on maximum achievable data rate in the underwater channel is investigated.
\end{abstract}

\section{Introduction}
Underwater optical wireless communications (UOWC) is an application in the field of optical wireless communications (OWC) that could complement the traditional accoustic method of transmission. Some of the inherent advantages of UOWC over the dominant acoustic underwater communications technology include lower latency and higher data rate~\cite{7450595}. These advantages, coupled with the lower transmission power associated with the semiconductor devices used in OWC, can enable remote high speed wireless communication over tens of metres. The removal of cables typically used for high speed communications could save time and money for ships in a harbour and for remotely operated vehicles (ROV) or for retrieving data from sensor nodes in coastal water or the open sea.

Designing an efficient optical wireless communications system for the underwater environment requires a detailed knowledge of the channel and an understanding of how these different types of water affect signal transmission. Previous studies in the literature have detailed methods for simulating the the absorption and scattering effects of the UOWC channel model~\cite{cox2012simulation,6413540,6685978,fletchprop,8886098} as well as the theory behind it~\cite{MobleyCurtisD1994Law:,mobley2001radiative}. Other studies have investigated turbulence in simulation as a separate entity~\cite{Vali:17}, and these results have been confirmed through experimental work~\cite{Vali:18}. However the photon tracking simulations in \cite{Vali:17,Vali:18} model turbulence in isolation from the absorbing and scattering components of the underwater channel and references~\cite{cox2012simulation,6413540,6685978,fletchprop,8886098} do not include turbulence in their modelling parameters. A more comprehensive approach to modelling the channel is to consider these three components in a single simulation as in our earlier work~\cite{8658289,Geld1907:Study}.

This paper describes a simulation that considers turbulence as a type of small angle scattering to improve the modelling of all the components of the UOWC channel. The main contributions of this paper are as follows: a comprehensive model is presented which accounts for turbulence as well as absorption and scattering; a comparison between the relative effects of turbulence in different water types in terms of temporal and spatial spreading is made; the distributions of received photon intensity for different conditions of water and turbulence are compared. 

The paper is structured as follows, Section~\ref{backgr} provides some background information on the theory, while Section~\ref{meth} outlines the methodology used in the channel simulation and Section~\ref{datameth} describes the simulation of data transmission through the channel. Results will be presented and their significance discussed in Section~\ref{res}, with final conclusions drawn in Section~\ref{conc}.

\section{Background}\label{backgr}
%In this section the background theory of the channel will be outlined in relation to absorption, scattering and turbulence. Additionally idea of turbulence induced scattering will be introduced and explained in greater detail.
\subsection{Absorption and Scattering}
When photons propagate through water, they are subject to scattering and absorption dependent upon the composition and condition of the water medium~\cite{7450595}. The probability of a single scattering or absorbing interaction as a photon travels through space is described by the absorption and scattering coefficients, denoted by $a(\lambda)$ and $b(\lambda)$ respectively, with $\lambda$ denoting the wavelength dependancy of the coefficients. The combined likelihood of any interaction taking place is therefore given by the extinction coefficient $c(\lambda)$ given by~\cite{cox2012simulation}:
\begin{equation}
c(\lambda)=a(\lambda)+b(\lambda).
\end{equation}
For simplicity throughout the rest of this paper when describing these coefficients $\lambda$ will be omitted from notation but it should be understood that these quantities are wavelength dependent. A higher $c$ means a photon is more likely to undergo a scattering or absorption event. Absorption occurs when a photon-particle interaction causes a photon to lose all of its energy and therefore stop propagating. Its effect on received signal power is entirely attenuating.  Scattering is of particular interest as it occurs when a photon-particle interaction results in a change of the photon's propagation path. This change in direction can be observed at the receiver (Rx) as attenuation but also as dispersion in both time and space due to multipath propagation~\cite{Geld1907:Study,fletchprop}. These coefficients have been evaluated through experimental studies, most notably by Petzold in~\cite{petzold1972volume}.%should i state why petzold's is "notable"?---explain how it can be seen as simply attenuation if the Rx is small and sample rate low???

The derivation of $a$ and $b$ from experimental measurements is outlined in~\cite{mobley,petzold1972volume}. The absorption coefficient, $a$, can be found by measuring the numbers of photons present at the Rx plane when a known number are transmitted through a small body of water. Conversely, the measurement of the scattering coefficient, $b$, requires the measurement of the volume scattering function (VSF), $\beta(\theta)$, which dictates the probability of a photon being scattered at a certain angle. This process involves moving an Rx around a point and measuring the number of photons incident at each angle of displacement. When $\beta(\theta)$ is integrated over all possible angles the likelihood of a single scattering interaction becomes~\cite{cox2012simulation}:%maybe change the bit about measurement of scattering?
\begin{equation}
b=2\pi\int_{0}^{\pi}{\beta(\theta)\sin(\theta)}d\theta.
\end{equation}

\subsection{Turbulence}
A third component of the UOWC channel is turbulence. Turbulence in water is caused by fluctuations in the refractive index which arise from random variations in salinity and temperature~\cite{MobleyCurtisD1994Law:}. When a photon propagates through the turbulent UOWC channel, these `pockets' with different refractive indices cause an alteration in direction. At the Rx this may be realised in the form of fluctuations in received power. This variation in power is described by the scintillation index, defined as~\cite{7450595}:
\begin{equation}
\sigma_I^2=\frac{\langle I^2 \rangle-\langle I \rangle^2}{\langle I \rangle^2},
\end{equation}
where $I$ is the received intensity and $\langle . \rangle$ denotes the ensemble average. It may be useful for researchers to have a closed form expression for $\sigma_I^2$ that takes all channel parameters into account, in the form:
\begin{equation}
\sigma_I^2=f(T, S ,a ,b ,Z_{link} , ...),
\label{imposs}
\end{equation}
where $f()$ is a function of temperature $T$; and salinity, $S$; as well as the transmission length, $Z_{link}$; absorption; and scattering coefficients, and any other channel parameter. However to the best of the authors knowledge no such expression exists in literature, although one for pure water, i.e. ignoring absorption and scattering parameters, was reported in~\cite{doi:10.1080/17455030.2012.656731}.

\subsection{Turbulence Induced Scattering}
If an expression that satisfies \eqref{imposs} exists then it cannot be found unless absorption, scattering, and turbulence are brought together in a single simulation. Previous research into underwater turbulence has treated absorption and scattering as distinct properties of the channel and used a log-normal fading model to account for turbulence at the Rx~\cite{7485523,7928991,8559096}. This method of modelling the fading statistics only ignores the temporal and spatial dispersion due to the turbulent channel. 
 
An alternative way of thinking about turbulence in simulation is to consider it as a small-angle scattering effect. This idea was discussed in \cite{MobleyCurtisD1994Law:} and developed through simulation in our earlier work~\cite{8658289,Geld1907:Study}. A similar technique was used in~\cite{Vali:17}, although their approach did not consider effects of scattering due to particles as well as turbulence.

The scattering coefficient can be split into its constituent parts to allow their contributions to be examined separately. Throughout this paper, subscripts $sw$, $p$, and $t$ will be used to refer to scattering contributions from seawater, particles, and turbulence respectively. Thus:
\begin{equation}
b=b_{sw}+b_{p}+b_{t}.
\label{eqn:b}
\end{equation}
When Petzold measured the commonly used values of $b$ in certain water types the accuracy was limited by a sensitivity of $0.1^o$~\cite{petzold1972volume}. Therefore, commonly used $b$ values do not include the smallest angle scattering - which in a turbulent channel is the most common. If there is an increased small angle probability in $\beta(\theta)$, then $b$ must be adjusted due to the relationship between the two parameters. It has been shown experimentally that underwater turbulence induces scattering at angles below this sensitivity~\cite{bogucki2004light}. This small angle scattering has been omitted from prior works due to the misconception that scattering at very small angles is analogous to no scattering at all as in~\cite{gordon1993sensitivity}. However if the link length is greater than $b^{-1}$ then the mean number of scattering interactions, $N_{scatter}=b L$, per channel will be greater than 1. As such multiple scattering must be considered.

Based on these findings and in order to account for turbulence induced scattering accurately, expression~\eqref{eqn:b} should be revised to show Petzold's value with the addition of an adjustment term. That is:
\begin{equation}
b=b_{Petzold}+b_{t}.
\end{equation}
As integration is a linear operation, the total VSF of any water channel can also be split into its constituent parts. Then it can be represented by the sum of each component's VSF weighted by its scattering coefficient. That is:
\begin{equation}
b\beta(\theta)=b_{sw}\beta_{sw}(\theta)+b_{p}\beta_{p}(\theta)+b_{t}\beta_{t}(\theta)
\label{beta total}.
\end{equation}
The VSF of seawater and particle induced scattering are represented by Mie scattering and a Henyey-Greenstein (HG) function respectively as in literature~\cite{mobley,7450595,MobleyCurtisD1994Law:}: 
\begin{equation}
\beta_{sw}(\theta)=0.06225(1+0.835\cos(\theta)^2)
\end{equation}
\begin{equation}
\beta_p(\theta)=\frac{1-g^2}{4\pi\left(1+g^2-2g\cos(\theta)\right)},
\end{equation}
for all water conditions $\beta_p(\theta)$ was modelled with an average cosine, $g$, equal to 0.975 in order to have a general VSF shape with only the number of interactions per channel changing. The newly introduced turbulent scattering is modelled through the Fournier-Forand (FF) function~\cite{fournier1994analytic}, this model links $\beta(\theta)$ to the refractive index of water and is highly weighted towards small angles. It is given as: 
\begin{equation}
\begin{split}
\beta_t(\theta) & =\frac{1}{4\pi(1-\delta)^2\delta^v}[v(1-\delta)\\ &-(1-\delta^v)+[\delta(1-\delta^v)\\ &-v(1-\delta)]\sin^{-2}\left(\frac{\theta}{2}\right)]\\ &+\frac{1-\delta^v_{180}}{16\pi(\delta^v_{180}-1)\delta^v_{180}}(3\cos^2\theta-1),
\end{split}
\end{equation}
where, $v=\frac{3-m}{2}$ with $m=3.05$ being the Junge slope parameter. Also $\delta$ is defined as:
\begin{equation}\label{delta}
\delta=\frac{4}{3}(n-1)^{-2}\sin\left(\frac{\theta}{2}\right)^2,
\end{equation}
where $n=1.33$ is the refractive index of water and $\delta^v_{180}$ is expression~\eqref{delta} evaluated at $\theta=180$\textsuperscript{o}.

\section{Channel Modelling}\label{meth}
This section describes the simulation methodology used in the modelling of a uniform multiple scattering channel. It later outlines the double-gamma function (DGF) commonly used in UOWC.
\subsection{Photon Tracking MC Simulation Model}
\begin{figure}[h]
	\centering
	\includegraphics[width=0.4\textwidth]{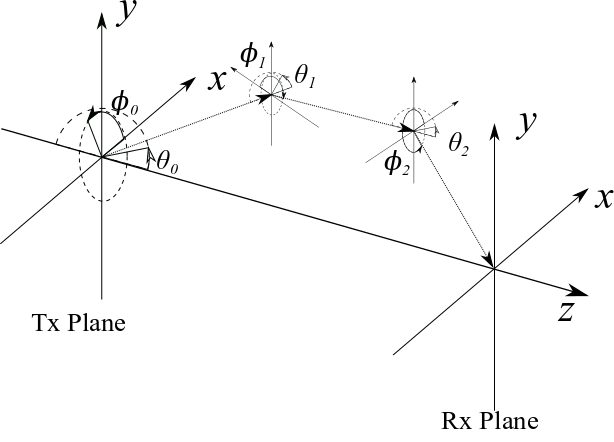}
	\caption{Graphic showing the tracking of a photon from Tx to Rx plane, propagating along the z-axis. The global x-y-z axis are shown as well as the local axis that are used to find the new direction of a scattered photon.}
	\label{track}
\end{figure}
The channel is simulated via a photon tracking MC method similar to that described in~\cite{cox2012simulation}. Petzold's values of $a=0.295$~m\textsuperscript{-1} and $b_{petzold}=1.875$~m\textsuperscript{-1} are used to model turbid harbour water with $a=0.179$~m\textsuperscript{-1} and $b_{petzold}=0.219$~m\textsuperscript{-1} for coastal water. In line with \eqref{eqn:b}, $b_{sw}=2.33\times~10^{-3}$~m\textsuperscript{-1} and $b_p=b_{Petzold}-b_{sw}$~m\textsuperscript{-1}. Initially the adjustment term, $b_t$, is chosen based on literature which suggests it can be as large as $10$~m\textsuperscript{-1} in some channels~\cite{bogucki2004light}. It is then fine tuned based on $\sigma_I^2$ estimates from simulation results to ensure weak turbulence is modelled. Photons are emitted at a uniformly distributed random angle within the transmitters (Tx) radiation angle, they then travel a random distance, $\Delta s$, dependant upon the extinction coefficient, $c$, before undergoing an interaction event. This distance is calculated by~\cite{cox2012simulation}:
\begin{equation}\label{distance}
\Delta s=-\frac{\ln \varepsilon_s}{c},
\end{equation}
where $\varepsilon_s$ is a uniformly distributed random number between 0 and 1.

All photons are assigned a weighting factor prior to emission. Initially $W_1=1$, and at each interaction this weight, $W_n$, is multiplied by the single scattering albedo to represent the effects of both scattering and absorption. This process is represented by~\eqref{weight}.
\begin{equation}\label{weight}
W_{n+1}=W_n\left(\frac{b}{c}\right).
\end{equation}
When $W_n$ falls below a defined threshold then the photon is discarded from the simulation. If the photon is not discarded then its new scattered direction must be updated in three dimensional space. The longitudinal angle from the z-axis, $\theta_s$, and radial angle, $\phi_s$ are such that:
\begin{equation}\label{theta}
%\left.\frac{d}{d\theta}\left(\frac{\varepsilon_\theta}{2\pi}\right)={\tilde{\beta}(\theta)\sin(\theta)}~~~~\middle|\right._{\theta=\theta_s}
\int_{0}^{\theta_s}{\tilde{\beta}(\theta)\sin(\theta)}d\theta=\varepsilon_\theta,
\end{equation}
\begin{equation}\label{phi}
\phi_s=2\pi\varepsilon_\phi,
\end{equation}
where $\varepsilon_\theta$ and $\varepsilon_\phi$ are uniformly distributed random numbers between 0 and 1 and $\tilde{\beta}(\theta)$ denotes the use of a normalised VSF. Solving expression~\eqref{theta} for $\theta_s$ for a given $\varepsilon_\theta$ provides the random scattering angle for a single interaction. After each scattering interaction, the photon's new position is calculated in a 3-dimensional coordinate plane as illustrated in Fig.~\ref{track}. Once its z-coordinate is greater than the link distance then the tracking iteration is stopped and its point of intersection with the Rx plane is noted along with its $W_n$ at that point. The sum of all $W_n$ values corresponding to a particular discrete coordinate block normalised by the total number of photons in the simulation gives the probability of a photon propagating through the channel reaching that point. When the arrival time is of interest then that can be calculated for each photon by taking the sum of all $\Delta s$ and dividing by the speed of light in water. These values can be mapped to discrete time blocks to find the likelihood of a photon arriving with a given time interval. This simulation approach is an accurate representation of a solution to the radiative transfer equations for the underwater channel~\cite{mobley}.

For our simulation we use a laser diode point source with radiation angle of 1.5~mrad and a threshold weighting of $10^{-6}$. Channel lengths of 15~m and 30~m are used to allow a comparison between different water types and link ranges. The Rx plane is made up of discrete (x,y) coordinate blocks of 10~cm. When the impulse response of the channel is to be investigated then the weighted received photons are put into a weighted histogram with time blocks of $10^{-10}$~s. These discrete blocks correspond to an Rx radius of 5~cm which allows for a sensitivity high enough for the temporal and spatial distributions of the channel response to be examined in detail.

\subsection{The Double-Gamma Channel Model}
The double-gamma function (DGF) was originally proposed for the impulse response modelling of photons travelling through clouds~\cite{mooradian1982temporal}. It uses two separate Gamma functions to represent the line of sight (LOS) and non line of sight (NLOS) components of the channel and is shown in equation~\eqref{dgf}. The non-scattered direct path is unlikely to exist in channels that have a high enough $b$ to cause multiple scattering. As such a good fit can be found for the impulse response from the Monte-Carlo (MC) simulation with only two gamma functions. The DGF is fitted to the impulse response generated MC to give a mathematical description of the time domain response of the channel. Additionally it also has the advantage of saving memory space in storing only the constants of a known function rather than the full output of the MC simulation. The temporal response of a channel described by the DGF is given by:
\begin{equation}\label{dgf}
h(t)=C_1t\exp(-C_2t)+C_3t\exp(-C_4t).
\end{equation}
Where $C_i$, $i={[1,2,3,4]}$ are constants found through least-squares curve fitting.
%\begin{equation}
%C_{i}=\text{argmin}\left(\int_{0}^{\infty}{\left(h(t)-h_{MC}(t)\right)^2}dt\right).
%\end{equation}

\section{Modelling of Data Transmission}\label{datameth}
The channel model described in Section~\ref{meth} is used in this section to model data transmission through the UOWC channels.
\subsection{Discrete Time Channel Model}\label{sec:dischan}
The result of the convolution of $h(t)$ and signal pulse shape, $f(t)$, shows the channels effect on a transmitted signal.
It is useful for further analysis to break up this received signal into discrete time blocks of duration $T_b$:
\begin{equation}\label{discrete}
p_k=\int_{(k-1)T_b}^{kT_b}f(t)*h(t)dt.
\end{equation}
These discrete blocks, $p_k$, represent the power in a single bit interval due to a transmitted pulse of duration $T_b$. In a channel with no bandwidth limitation then all the power would be confined to $p_1$, however when this is not the case some power leaks into neighbouring time slots resulting in inter-symbol interference (ISI).

This discrete-time block model of a signal allows the simulation of the effects on a transmitted bit stream due to ISI caused by the channel impulse response. The number of photons incident on an Rx follows a random Poisson distribution with mean $a_i$~\cite{fletchprop}. The mean photon rate in the $i^{th}$ bit slot is dependent upon the discrete channel response described by~\eqref{discrete} and the transmitted bits, $x_{i-k}$, such that:
\begin{equation}\label{phot_chan}
a_i=N_{ph}\left(x_i p_1 + \sum_{k=2}^{M}{x_{i-k} p_k}\right)+n_{bg},
\end{equation}
where $k$ denotes the interfering bit up to the maximum channel memory $M$. Furthermore, $N_{ph}$ is the mean number of photons emitted during a bit `1' and $n_{bg}$ is the mean photon count from background noise sources. The mean value of $n_{bg}$ is estimated for typical water conditions using expressions in~\cite{1178461}. Whilst $N_{ph}$ is found via~\cite{Dong:2013:CUW:2532378.2532391}:
\begin{equation}
N_{ph}=\frac{2P_tT_b\lambda}{hv},
\end{equation}
where $P_t$ is the average transmit power, $T_b$ is the bit duration, $\lambda$ is the wavelength of the source, $h$ is Planck's constant, and $v$ is the speed of light.

\subsection{Maximum Achievable Data Rate}\label{sec:maxrate}
Here we investigate the effect of underwater turbulence on the maximum achievable data rate with on-off keying (OOK) as a case study. The maximum data rate supported by a channel for OOK is equal to the greatest bit rate for which the mutual information, $I(X;Y)$, is equal to 1. This means that the channel input, $X$, yields a proportional change in output, $Y$. When $I(X;Y)<1$ then changes in $X$ no longer yield a proportional $Y$, so the output does not change with the input. In this case, instantaneous output $y_i$ is proportional to both the current and previous inputs $x_{i-k}$ where $k=1, 2,..., M$ meaning the ISI component of the channel is not insignificant. When this is the case then the data rate no longer increases with the symbol rate, $R_{sym}$, as the information carried in each transmitted symbol is no longer 1 bit. This means that as the symbol rate continues to increase, the data rate remains limited to a maximum rate defined by the characteristics of the channel. Mathematically the capacity of a channel in bits per symbol can be written as~\cite{490551}:
\begin{equation}\label{max_rate}
C=\lim\limits_{L\to\infty}\max_{P(X)}(I(X;Y)),
\end{equation}
here $P(X)$ denotes that $I(X;Y)$ is evaluated over all possible input distributions for an L bit input stream. The maximum data rate in bits per second supported by a channel is simply:
\begin{equation}\label{Rmax}
R_{max}=C\times R_{sym}.
\end{equation}

The mutual information of a channel can be calculated as the difference between the entropy rates of the output, $H(Y)$, and channel, $H(Y|X)$~\cite{6392505,Dong:2013:CUW:2532378.2532391}:%maybe use different citations here?
\begin{equation}\label{mutual}
I(X;Y)=H(Y)-H(Y|X).
\end{equation}
When a random bit stream is transmitted through a channel, the received signal will be stochastic due to the effects of the channel and the random signal. %unsure about this sentence but it does describe whats happening
The respective output and channel entropies are estimated by the following relationships:
\begin{equation}
-\frac{1}{L}\log_2(P(Y^L)) \to H(Y)~~~~~~~~~~~~L \to \infty
\end{equation}
\begin{equation}\label{HYX}
-\frac{1}{L}\log_2(P(Y^L|X^L)) \to H(Y|X)~~~L \to \infty
\end{equation}
The channel entropy, $H(Y|X)$, is calculated based on $P(Y^L|X^L)$ which is the probability of an $L$ bit output stream, $Y$, given an $L$ bit input stream $X$. This can be simplified for calculations at each timeslot as:
\begin{equation}
P(Y^L|X^L)=\prod_{i=1}^{L}{P(y_i|x_i)}.
\end{equation}
This can be simplified further through the relationship between $x_i$ and the channel described by~\eqref{phot_chan} so:
\begin{equation}
{P(y_i|x_i)}={P(y_i|a_i)}.
\end{equation}
Here, the probability of a given photon count, $y_i$, for a mean photon arrival rate, $a_i$, is described by a Poisson distribution such that~\cite{fletchprop}:
\begin{equation}
{P(y_i|a_i)}=\frac{{a_i}^{y_i}}{y_i!}\exp{(-a_i)}.
\end{equation}

The estimation of $P(Y^L)$ utilises the discrete channel model and Bahl, Cocke, Jelinek, and Raviv (BCJR) algorithm, as described in~\cite{6392505,Dong:2013:CUW:2532378.2532391}. Without knowledge of the transmitted data stream it is evident from~\eqref{discrete}~and~\eqref{phot_chan} that there are $2^M$ possible states, $s_i$, the channel can enter. This is dependant upon the previous $M$ input bits, that can give rise to $y_i$. By this relationship we get:%is this added part good? re-read tomorrow to check
\begin{equation}
P(Y^L)=\prod_{i=1}^{L}P(y_i),
\end{equation}
where,
\begin{equation}
P(y_i)=\sum_{s_i}\sum_{s_{i-1}} p(y_i|s_i,s_{i-1}).
\end{equation}
The channel can only enter a new state, $s_{i}$, if the state transition is possible from any channel input. As such the state transition probability can be defined as:
\begin{equation}
\gamma_i(s_i,s_{i-1})=p(x)q(s_i,s_{i-1})p(y|s_i,s_{i-1}),
\end{equation}
where $p(x)$ is the probability of the input changing (which is equal to $0.5$ for OOK), $q(s_i,s_{i-1})$ is a transition function that is equal to 1 if a state transition $s_{i-1} \to s_i$ is possible and 0 otherwise. Using the forward recursion part of the BCJR algorithm this can become something of similar form to $P(Y^L)$,
\begin{equation}\label{alpha}
\alpha_{i+1}(s_i)=\sum_{s_{i-1}}\alpha_{i}(s_{i-1})\gamma_i(s_i,s_{i-1}),
\end{equation}
therefore:
\begin{equation}
P(Y^L)=\sum_{s_i}\alpha_{L}(s_i).
\end{equation}

However, due to the recursive multiplication in~\eqref{alpha}, $\alpha_i(s_i)$ tends to 0 very quickly for a large $i$. Resultantly, for computation it is useful to normalise at each bit slot, as in:
\begin{equation}
A_i=\frac{1}{\sum_{s_i}{\alpha_i(s_i)}}.
\end{equation}

Finally, $H(Y)$ can be written in terms of $A_i$ such that:
\begin{equation}\label{HY}
\frac{1}{L}\sum_{i=1}^{L}\log_2{(A_i)} \to H(Y)~~ L \to \infty.
\end{equation}
Equations~\eqref{HYX} and~\eqref{HY} can then be substituted into~\eqref{mutual} and~\eqref{max_rate} in order to estimate the maximum data rate supported by the channel.

\section{Numerical Simulation, Results and Discussion}\label{res}
Detailed in this section are the results of simulation processes described in Sections~\ref{meth}~and~\ref{datameth}.
\subsection{Temporal and Spatial Distributions}
\subsubsection{Photon Spatial Distribution}
\begin{figure}[h]
	\centering\includegraphics[width=0.45\textwidth]{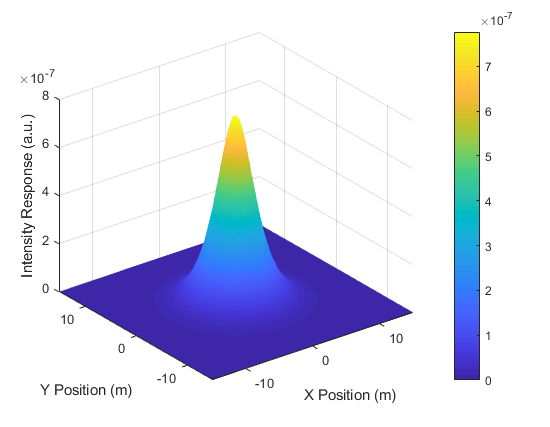}
	\caption{Rx plane spatial intensity distribution for harbour water with no turbulence.}\label{space0}
\end{figure}
The photon tracking MC simulation is run for various channel configurations as described in Section~\ref{meth} for $10^9$ photons with an Rx field of view of 10\textsuperscript{o}. The received intensity distribution for the harbour water channel, when $b_t=0$~m\textsuperscript{-1} (i.e. no turbulence induced scattering), is shown in Fig.~\ref{space0}. This can be seen to have a noticeably symmetrical Gaussian shape centred around the (0,0) coordinates in the Rx plane. The symmetrical shape is in line with the expected due to the radial scattering angles, calculated with~\eqref{phi}, being uniformly distributed over all angles. The rest of this section will therefore use 1-dimensional plots to make for a simpler comparison between the various channel conditions considered.
%\begin{figure}[h]
	%\centering\includegraphics[width=0.45\textwidth]{coast30_1d}
	%\caption{Rx plane spatial intensity distribution for coastal water on the positive x-axis for different conditions of turbulence induced %scattering for a link distance of 30~m.}\label{spacecoast}
%\end{figure}
%\begin{figure}[h]
	%\centering\includegraphics[width=0.45\textwidth]{1harbour15_1d}
	%\caption{Rx plane spatial intensity distribution for harbour water on the positive x-axis for different conditions of turbulence induced scattering for a link distance of 15~m.}\label{spaceharbour}
%\end{figure}

\begin{figure}[h]
	%\centering\includegraphics[width=0.45\textwidth]{1d_coast_and_harb_anotated}
\centering\includegraphics[width=0.45\textwidth]{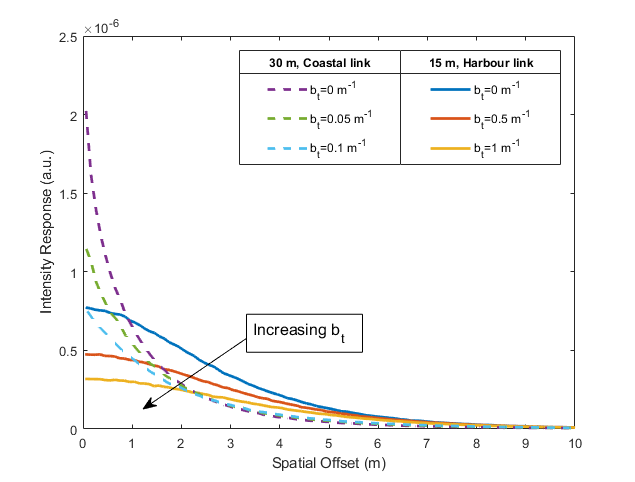}
	\caption{Rx plane spatial intensity distribution for 30~m coastal, and 15~m harbour water link with different conditions of turbulence induced scattering.}\label{spaceboth}
\end{figure}
Shown in Fig.~\ref{spaceboth} are the 2-dimensional Rx plane intensity distributions in both coastal and harbour waters at different values of $b_t$. When comparing the spatial distributions for $b_t=0$~m\textsuperscript{-1} it is noticeable that for the case of highly scattering harbour water the intensity is scattered over a much greater area in the Rx plane. In fact the full width half maximum (FWHM) beamwidth is approximately $5.5$~m whereas for coastal water it is $1.1$~m. These figures are in line with the expectation of more photon-particle interactions due to the longer extinction length, $cZ_{link}$, of the harbour link. The extinction length in a $15$~m harbour channel is $33$ interactions, compared to $11$ for a coastal channel of $L=30$~m. For both water types this FWHM beamwidth increases as $b_t$ increases, however the magnitude of its contribution is different for each water type, expectantly. 
%An interesting feature across the two channels is the difference in the effect of the addition of turbulence induced scattering. For the case of highly scattering harbour water in Fig.~\ref{spaceboth}, the decrease in aligned intensity with no offset is $7.7\times10^{-7}$ when $b_t=0$~m\textsuperscript{-1} compared to $4.7\times10^{-7}$ and $3.2\times10^{-7}$ with $b_t$ equal to $0.5$ and $1$~m\textsuperscript{-1} respectively. However when considering coastal water, with its lower $b_{Petzold}$ value, similar percentage decreases are observed for lower changes in $b_t$. In this case with $b_t=0$~m\textsuperscript{-1} the intensity is $3.5\times10^{-6}$ which then decreases to $1.2\times10^{-6}$ for $b_t=0.05$~m\textsuperscript{-1} and $8\times10^{-7}$ for $b_t=0.1$~m\textsuperscript{-1} - representing decreases to 34\% and 23\% of the case with no turbulence this despite the smaller $b_t$ values used. This implies that the impact of turbulence induced scattering depends on water type. The shape too is different, the intensity response being much more concentrated near 0~m spatial offset in the 30~m harbour link.

An interesting feature across the two channels is the difference in the relative impact of turbulence induced scattering. For the case of the highly scattering 15~m harbour channel in Fig.~\ref{spaceboth}, the channel gain at perfect alignment (i.e. Spatial Offset = 0~m) is $7.7\times10^{-7}$ when $b_t=0$~m\textsuperscript{-1} and it decreases by 58\% when $b_t$ increases to 1~m\textsuperscript{-1}. Comparatively, in the 30~m coastal channel the aligned channel gain for $b_t=0$ is $3.5\times10^{-6}$ and it falls by 65\% when $b_t=0.05$~m\textsuperscript{-1}, and 77\% when $b_t=0.1$~m\textsuperscript{-1}. This can be explained by returning to the definition of turbulence induced scattering as being a case of small angle scattering. When the number of scattering interactions per channel is high then the photons are already dispersive and therefore have some resilience to additional scattering at small angles. However, when the number of interactions per channel is lower, as is the case with the 30~m coastal link, the increase can have a much greater impact on the propagation path taken by a photon. Meaning that proportionally turbulence induced scattering has a lesser effect in a channel that already has a high probability of scattering compared to one that does not.
%A further possible explanation is that with no turbulence, harbour water has a single scattering albedo ($\omega_0$) of $0.88$, meaning when a photon interacts with a particle there is an 88\% chance of it being scattered rather than absorbed, for coastal water $\omega_0=0.55$. Combined with the increased number of photon-particle interactions in harbour water this means that not only are there more photon-particle interactions in the harbour water case, but the likelihood of these interactions producing scattering is also higher. This effect yields the wider distribution and FWHM beamwidth. However, when $b_t$ is added to the simulation its relative impact can be quantified by how it changes the $\omega_0$ and $cZ_{link}$ of the channels. For coastal water when $b_t=0.01$~m\textsuperscript{-1}, $\omega_0$ increases to $0.60$ and then $0.64$ when  $b_t=0.1$~m\textsuperscript{-1} whilst $cZ_{link}$ becomes $13.4$ and $15$. Meanwhile for the $15$~m harbour link $\omega_0$ rises to $0.89$ and $0.91$ for $b_t=0.5$ and $1$~m\textsuperscript{-1} respectively, and $cZ_{link}$ becomes $41$ and $48$ for the same conditions. It is apparent from these values that proportionally turbulence induced scattering has a lesser effect in a channel that already has a high probability of scattering compared to one that does not.

\subsubsection{Impulse Response}
%\begin{figure}
%	\begin{subfigure}[h]{0.45\textwidth}
%	\includegraphics[width=\textwidth]{dgf_fit_coast30bt0_smooth}
%	\caption{30~m coastal water link.}
%	\label{coastir}
%\end{subfigure}
%\begin{subfigure}[h]{0.45\textwidth}
%	\includegraphics[width=\textwidth]{dgf_fit_harbour15bt0_smooth}
%	\caption{15~m harbour water link.}
%	\label{harbourir}
%\end{subfigure}
%	\caption{Impulse response of a 30~m coastal water link and 15~m harbour water link, with $b_t=~0$~m\textsuperscript{-1}, from MC simulation with fitted DGF.}
%	\label{ir_both}
%\end{figure}
\begin{figure}
\centering
	\includegraphics[width=0.45\textwidth]{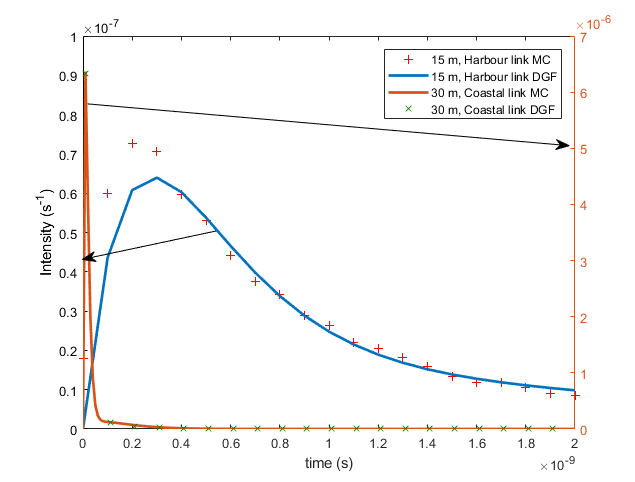}
	\caption{Impulse response of a 30~m coastal water link (right y-axis) and 15~m harbour water link (left y-axis), with $b_t=~0$~m\textsuperscript{-1}, from MC simulation with fitted DGF.}
	\label{ir_both}
\end{figure}
Impulse responses generated through the previously described MC simulation are fitted to a DGF in order to get an expression for the channel that takes temporal characteristics into account. Fig.~\ref{ir_both} shows the DGF fitted to the simulation data for the cases of no turbulence induced scattering in coastal and harbour links respectively. It is evident that the 15~m harbour link is far more dispersive than the 30~m coastal link. 

\begin{figure}
	\centering
	\begin{subfigure}[h]{0.45\textwidth}
		\includegraphics[width=\textwidth]{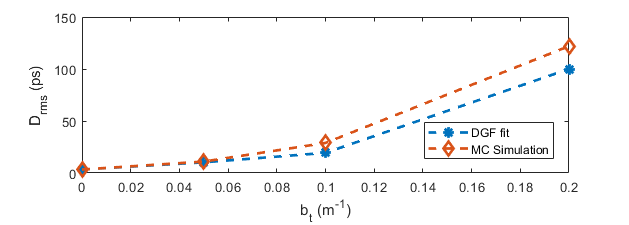}
		\caption{30~m coastal water link.}
		\label{coast_Dfit}
	\end{subfigure}
	\\	
	\begin{subfigure}[h]{0.45\textwidth}
		\includegraphics[width=\textwidth]{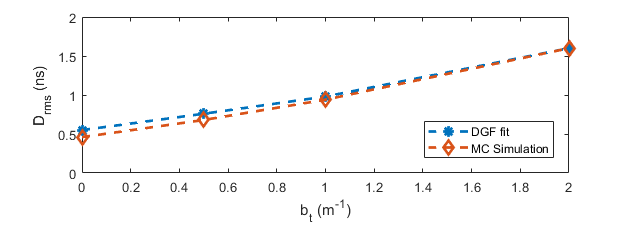}
		\caption{15~m harbour water link.}
		\label{harbour_Dfit}
	\end{subfigure}
	\caption{$D_{rms}$ vs $b_t$ for different UOWC links highlighting the relative impact of turbulence induced scattering on temporal dispersion.}
	\label{Dfit}
\end{figure}

The DGF fits are compared to the MC responses in two ways, the $R^2$ - metric which is a measure of fit~\cite{montgomery2007applied} - and the root mean squared delay spread ($D_{rms}$). The $D_{rms}$ of the simulation is denoted by $D_{rms}^{sim}$, whilst that of the DGF fit is $D_{rms}^{DGF}$. Both $D_{rms}^{sim}$ and $D_{rms}^{DGF}$ are plotted against $b_t$ in Fig.~\ref{Dfit} and are shown to be close for all turbulence conditions simulated. The DGF coefficients for channel conditions used later in this paper are presented along with the respective $R^2$ in Table~\ref{Cfittab}. The $R^2$ value for each fit is greater than 0.9 which indicates the DGF model describes MC data set adequately for all parameters used. This justifies the use of the DGF fits rather than raw simulation data in the rest of the paper.
\begin{table}[h]
	\begin{center}
		\caption{Fit parameters for DGF for channel conditions used in this paper}
		\begin{tabular}{| l | l | l | l | l | l |}
			\hline
			$b_{t}$~(m\textsuperscript{-1}) & $C_1$ & $C_2$ & $C_3$ & $C_4$ & $R^2$\\ \hline
			\multicolumn{6}{|c|}{30~m, Coastal Water} \\ \hline
			0 & $2\times10^{6}$ & $1.15\times10^{11}$ & 4072 & $1.2\times10^{10}$ & 1.00 \\ \hline
			0.05 & $5.5\times10^{5}$ & $1\times10^{11}$ & 2544 & $1\times10^{10}$ & 1.00 \\ \hline
			0.1 & $1.86\times10^{5}$ & $9\times10^{10}$ & 2953 & $1.2\times10^{10}$ & 1.00 \\ \hline
			0.2 & $1.2\times10^{4}$ & $3\times10^{10}$ & 435 &$6\times10^{9}$ & 0.90 \\ \hline
			\multicolumn{6}{|c|}{15~m, Harbour Water} \\ \hline
			0 & 600 & $3.8\times10^{9}$ & 28.49 & $9\times10^{8}$ & 0.97 \\ \hline
			0.5 & 160 & $3.0\times10^{9}$ & 23.69 & $9\times10^{8}$ & 0.99 \\ \hline
			1 & 60 & $2.4\times10^{9}$ & 12.51 & $7.5\times10^{8}$ & 0.97 \\ \hline
			2 & 8.9 & $1.2\times10^{9}$ & 2.40 & $4.5\times10^{8}$ & 0.95\\ \hline
			%5 & 0.2 & $2.5\times10^{8}$ & 0.06 & $2\times10^{8}$ & 0.85 \\ \hline
		\end{tabular}
		\label{Cfittab}
	\end{center}
\end{table}
 %shown to be very close to $D_{rms}^{DGF}$ showing that the DGF provides a good fit in the time domain and confirms that the fit can be used in further investigations to accurately model the channel. All the fits used in Fig.~\ref{Dfit} have $R^2>0.9$ showing that DGF provides a very good fit for the impulse response MC data.
%The $R^2$ metric provides a further measure of how well a function describes a set of data. The closer this value is to 1 the better the fit. Therefore, it is apparent from the results shown in Table~\ref{IRtab} that DGF provides a very good fit for the impulse response MC data. This additionally justifies the use of the DGF in place of the MC impulse response data as a  description of the impulse response of UOWC links that allows for mathematical analysis with respect to communication parameters.
%The $R^2$ metric provides a further measure of how well a function describes a set of data. The closer this value is to 1 the better the fit. All the fits used in Fig.~\ref{Dfit} have $R^2>0.9$ showing that DGF provides a very good fit for the impulse response MC data. This additionally justifies the use of the DGF in place of the MC impulse response data as a  description of the impulse response of UOWC links that allows for mathematical analysis with respect to communication parameters.

Fig.~\ref{coast_Dfit}~and~\ref{harbour_Dfit} show $D_{rms}$ against $b_t$ for a 30~m coastal link and 15~m harbour link respectively. This highlights the difference in the impact of turbulence induced scattering on $D_{rms}^{DGF}$ in the already highly scattering harbour water compared to the coastal water link. The values of $b_t$ used in Fig.~\ref{coast_Dfit} are 10 times smaller than those in Fig.~\ref{harbour_Dfit} however the increase in $D_{rms}^{DGF}$ compared to when $b_t=0$~m\textsuperscript{-1} is larger. For the case of $b_t=0.1$~m\textsuperscript{-1} in coastal water $D_{rms}^{DGF}$ is 5.81 times larger than $b_t=0$~m\textsuperscript{-1}. Whereas for $b_t=1$~m\textsuperscript{-1} in harbour water the difference is only 1.78 times. This again implies that a channel that already has a high level of multiple scattering possesses a higher level of tolerance to turbulence induced scattering. From a communications perspective this increase in temporal dispersion will cause the energy of a transmitted symbol to spread out leading to ISI in the received signal.
%\begin{table}
%	\caption{Simulation and DGF RMS delay spread and $R^2$ fit metric for a 30~m coastal and 15~m harbour water links with different $b_t$ values.}
%	\begin{center}
%		\begin{tabular}{| l | l | l | l |}
%			\hline
%			\multicolumn{4}{|c|}{30~m, Coastal Water} \\ \hline
%			$b_t$~m\textsuperscript{-1} & $D_{rms}^{sim}$~(ps) & $D_{rms}^{DGF}$~(ps) & $R^2$ \\ \hline
%			0 & 3.34 & 3.33 & 1.00 \\ \hline
%			0.05 & 11.01 & 10.25 & 1.00 \\ \hline
%			0.1 & 28.89 & 19.38 & 1.00 \\ \hline
%			0.2 & 122 & 100 & 0.90 \\ \hline
%			\multicolumn{4}{|c|}{15~m, Harbour Water} \\ \hline
%			$b_t$~m\textsuperscript{-1} & $D_{rms}^{sim}$~(ns) & $D_{rms}^{DGF}$~(ns) & $R^2$ \\ \hline
%			0 & 0.46 & 0.55 & 0.97 \\ \hline
%			0.5 & 0.68 & 0.76 & 0.99 \\ \hline
%			1 & 0.94 & 0.98 & 0.97 \\ \hline
%			2 & 1.60 & 1.60 & 0.95 \\ \hline
%		\end{tabular}
%		\label{IRtab}
%	\end{center}
%\end{table}
\subsection{Scintillation Index Estimation}
\begin{figure*}[!htbp]
	\centering
	\begin{subfigure}[h]{0.45\textwidth}
		\includegraphics[width=\textwidth]{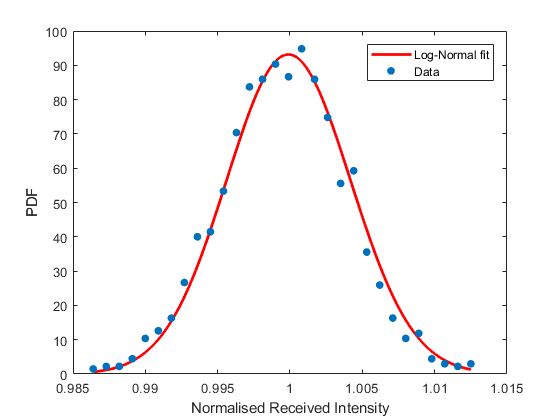}
		\caption{$b_{t_{max}}=0$~m\textsuperscript{-1}}
		\label{fig:cpdf0}
	\end{subfigure}
	~	
%	\begin{subfigure}[h]{0.45\textwidth}
%		\includegraphics[width=\textwidth]{pdffits_forjournal/pdfcoast15bt01_poorfit}
%		\caption{$b_{t_{max}}=0.1$~m\textsuperscript{-1}}
%		\label{fig:cpdf01}
%	\end{subfigure}
%	\\
%		\begin{subfigure}[h]{0.45\textwidth}
%			\includegraphics[width=\textwidth]{pdffits_forjournal/pdfcoast15bt02}
%			\caption{$b_{t_{max}}=0.2$~m\textsuperscript{-1}}
%			\label{fig:cpdf05}
%		\end{subfigure}
%		~	
		\begin{subfigure}[h]{0.45\textwidth}
			\includegraphics[width=\textwidth]{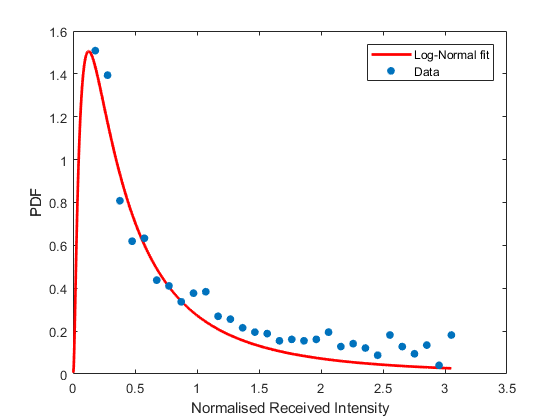}
			\caption{$b_{t_{max}}=0.3$~m\textsuperscript{-1}}
			\label{fig:cpdf1}
		\end{subfigure}
	\caption{Log-Normal pdf fits on histograms showing variation in received photon counts at the Rx of a 15~m coastal link.}
	\label{coast15pdf}
\end{figure*}

\begin{figure*}[!htbp]
	\centering
	\begin{subfigure}[h]{0.45\textwidth}
		\includegraphics[width=\textwidth]{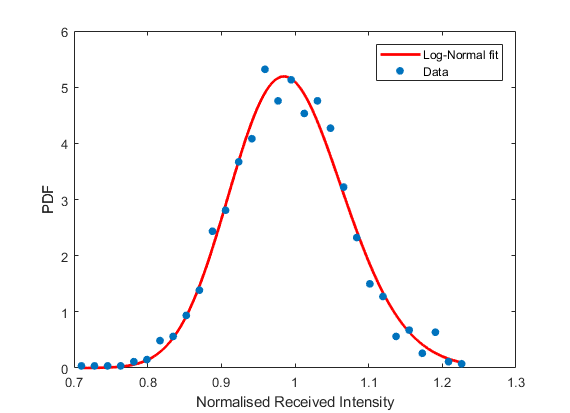}
		\caption{$b_{t_{max}}=0$~m\textsuperscript{-1}}
		\label{fig:cpdf030}
	\end{subfigure}
	~	
%	\begin{subfigure}[h]{0.45\textwidth}
%		\includegraphics[width=\textwidth]{pdffits_forjournal/pdfcoast30bt001}
%		\caption{$b_{t_{max}}=0.01$~m\textsuperscript{-1}}
%		\label{fig:cpdf0130}
%	\end{subfigure}
%	\\
%	\begin{subfigure}[h]{0.45\textwidth}
%		\includegraphics[width=\textwidth]{pdffits_forjournal/pdfcoast30bt005}
%		\caption{$b_{t_{max}}=0.05$~m\textsuperscript{-1}}
%		\label{fig:cpdf0530}
%	\end{subfigure}
%	~	
	\begin{subfigure}[h]{0.45\textwidth}
		\includegraphics[width=\textwidth]{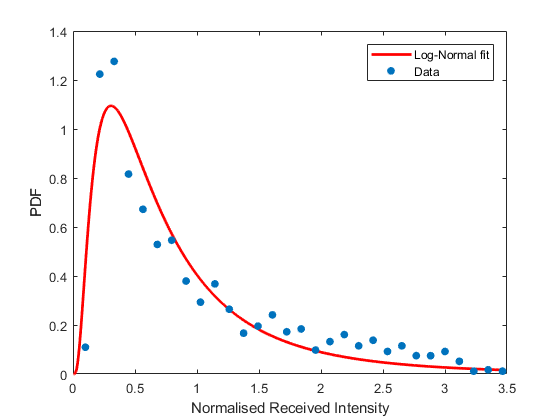}
		\caption{$b_{t_{max}}=0.15$~m\textsuperscript{-1}}
		\label{fig:cpdf130}
	\end{subfigure}
	\caption{Log-Normal pdf fits on histograms showing variation in received photon counts at the Rx of a 30~m coastal link.}
	\label{coast30pdf}
\end{figure*}
\begin{figure*}[!htbp]
	\centering
	\begin{subfigure}[h]{0.45\textwidth}
		\includegraphics[width=\textwidth]{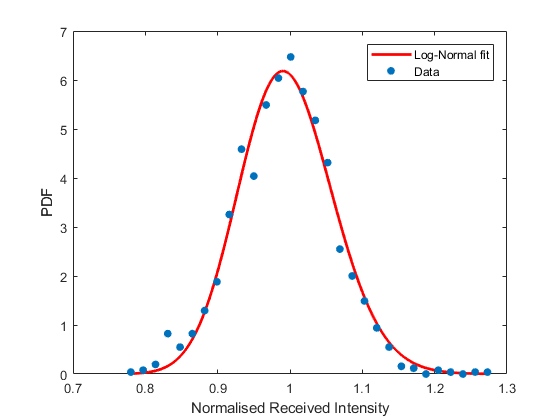}
		\caption{$b_{t_{max}}=0$~m\textsuperscript{-1}}
		\label{fig:hpdf0}
	\end{subfigure}
	~	
%	\begin{subfigure}[h]{0.45\textwidth}
%		\includegraphics[width=\textwidth]{pdffits_forjournal/pdfharbour15bt05}
%		\caption{$b_{t_{max}}=0.5$~m\textsuperscript{-1}}
%		\label{fig:hpdf05}
%	\end{subfigure}
%	\\
%	\begin{subfigure}[h]{0.45\textwidth}
%		\includegraphics[width=\textwidth]{pdffits_forjournal/pdfharbour15bt1}
%		\caption{$b_{t_{max}}=1$~m\textsuperscript{-1}}
%		\label{fig:hpdf1}
%	\end{subfigure}
%	~	
	\begin{subfigure}[h]{0.45\textwidth}
		\includegraphics[width=\textwidth]{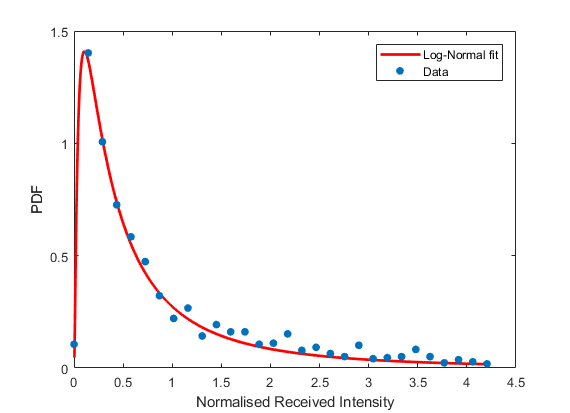}
		\caption{$b_{t_{max}}=5$~m\textsuperscript{-1}}
		\label{fig:hpdf5}
	\end{subfigure}
	\caption{Log-Normal pdf fits on histograms showing variation in received photon counts at the Rx of a 15~m harbour link.}
	\label{harbourpdf}
\end{figure*}
The random effect of $b_t$ on the received photon intensity is investigated by transmitting $10^7$ photons through the channel, repeated for $1500$ Monte-Carlo iterations, with $b_t$ randomly distributed on the interval $[0,b_{tmax}]$ each time. This is to represent the time varying nature of the underwater channel. This method of simulation allows each iteration of $10^7$ photons to be modelled as a uniform stationary channel, and is based on the assumption that the propagation time is shorter than the coherence time of the channel. The weighted sum of all photons received for each iteration are plotted as a histogram. The histogram is normalised to obtain a probability density function (pdf), and a Log-Normal function is fitted to this. The histograms and their respective fits are shown in Fig.~\ref{coast15pdf}~-~\ref{harbourpdf}. The mean normalised variance of the simulation and fitted Log-Normal scintillation index, $\sigma^2_{sim}$ and $\sigma^2_I$ respectively, as well as the mean log intensity, $\mu$, for these fits are given in Table~\ref{sitab}. The Log-Normal model of turbulence gives a reasonably good fit for the distributions with $R^2>0.75$ for all cases.
\begin{table}[h]
	\begin{center}
		\caption{Simulation and Log-Normal fit parameters for simulation of a turbulent coastal channel with link rages of 15 and 30~m.}
		\begin{tabular}{| l | l | l | l | l |}
			\hline
			$b_{tmax}$~(m\textsuperscript{-1}) & $\sigma_{sim}^2$ & $\sigma_I^2$ & $\mu$ & $R^2$ \\ \hline
			\multicolumn{5}{|c|}{15~m, Coastal Water} \\ \hline
			0 & $1.84\times10^{-5}$ & $1.84\times10^{-5}$ & $-9.21\times10^{-5}$ & 0.98 \\ \hline
		%	0.1 & 0.0840 & 0.1518 & -0.1210 & 0.37 \\ \hline
			0.2 & 0.3202 & 0.5297 & -0.3312 & 0.80 \\ \hline
			0.3 & 0.6399 & 0.8958 & -0.7696 & 0.90 \\ \hline
			\multicolumn{5}{|c|}{30~m, Coastal Water} \\ \hline
			0 & 0.0060 & 0.0061 & 0.0091 & 0.98 \\ \hline
			0.01 & 0.0105 & 0.0113 & -0.0110 & 0.98 \\ \hline
			0.05 & 0.1001 & 0.1437 & -0.0887 & 0.87 \\ \hline
			0.1 & 0.3204 & 0.4273 & -0.2671 & 0.83 \\ \hline
			0.15 & 0.5828 & 0.6999 & -0.4810 & 0.88 \\ \hline
			0.2 & 1.0374 & 2.0862 & -0.9923 & 0.82 \\ \hline
			\multicolumn{5}{|c|}{15~m, Harbour Water} \\ \hline
			0 & 0.0044 & 0.00042 & -0.0055 & 0.98 \\ \hline
			0.5 & 0.0271 & 0.0362 & -0.0295 & 0.89 \\ \hline
			1 & 0.0792 & 0.1087 & -0.0846 & 0.82 \\ \hline
			2 & 0.2389 & 0.3279 & -0.2238 & 0.79 \\ \hline
			5 & 0.8763 & 1.5284 & -0.7106 & 0.98 \\ \hline
		\end{tabular}
		\label{sitab}
	\end{center}
\end{table}

When $b_t=0$~m\textsuperscript{-1}, i.e. no turbulence, the intensity distributions across all three links have a Gaussian shape with normalised mean of approximately 1.  In Table~\ref{bt0gaus} the $R^2$ values for all three distributions is shown to be very close to 1. Indicating that the Gaussian distribution provides a very good fit, as would be expected for a link with no turbulence.

\begin{table}[h]
	\begin{center}
		\caption{Gaussian fit parameters for the three channel conditions with no turbulence.}
		\begin{tabular}{| l | l | l | l |}
			\hline
			~ & $\sigma^2$ & mean &$R^2$ \\ \hline
			Coast 15~m & $1.83\times10^{-5}$ & 0.9996 & 0.99 \\ \hline
			Coast 30~m & 0.0059 & 0.9896 & 0.98 \\ \hline
			Harbour 15~m & 0.0042 & 0.9933 & 0.98 \\ \hline
		\end{tabular}
		\label{bt0gaus}
	\end{center}
\end{table}

\begin{figure}
\centering
	\includegraphics[width=0.45\textwidth]{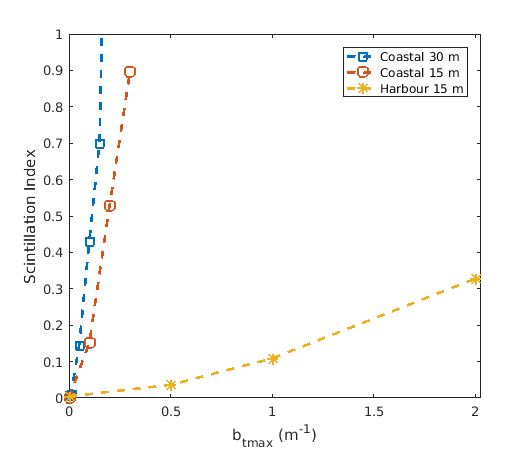}
	\caption{Log-Normal $\sigma_I^2$ vs $b_{tmax}$ in different water types and link ranges.}
	\label{si_vs_bt}
\end{figure}
In order to compare relative impact of turbulence induced scattering for different UOWC channels, $\sigma_I^2$ is plotted against $b_{tmax}$ in Fig.~\ref{si_vs_bt}. When comparing between channel configurations it is apparent that the relative impact of turbulence induced scattering is lessened when there are already a high number of scattering interactions per channel. In Fig.~\ref{si_vs_bt} it is clear that increasing the link distance for coastal water yields an increase in $\sigma_I^2$ for the same $b_{tmax}$. For example when $b_{tmax}=0.1$~m\textsuperscript{-1}, $\sigma_I^2=0.1518$ for a link distance of 15~m compared to 0.4273 when $Z_{link}=30$~m. This is in line with the $Z_{Link}$ dependency of the expression derived in reference~\cite{doi:10.1080/17455030.2012.656731}. However it is when comparing between the 15~m coastal and harbour links that the results deviate from when absorption and scattering are omitted. In the harbour channel a $b_{tmax}$ of 1~m\textsuperscript{-1} is required to yield a $\sigma_I^2>0.1$ compared to $b_{tmax}=0.1$~m\textsuperscript{-1} in coastal water over the same link range. This further adds to the suggestion that due to the multiple scattering nature of the UOWC channel, it is not valid to simply adapt concepts from free space optics (FSO) to the underwater channel.
\subsection{Maximum Achievable Rate Estimation}
%\begin{figure}
%	\centering
%\begin{subfigure}[h]{0.45\textwidth}
%	\includegraphics[width=\textwidth]{rate_vs_bt_coast30}
%	\caption{30~m coastal link.}
%	\label{capcoa}
%\end{subfigure}
%\\
%\begin{subfigure}[h]{0.45\textwidth}
%	\includegraphics[width=\textwidth]{rate_vs_bt_harbour15}
%	\caption{15~m harbour link.}
%	\label{caphar}
%\end{subfigure}
%\caption{Maximum achievable data rate against $b_t$ with a $20$~mW optical transmit power.}
%\label{capacity}
%\end{figure}
\begin{figure}[h]
\centering
	\includegraphics[width=0.45\textwidth]{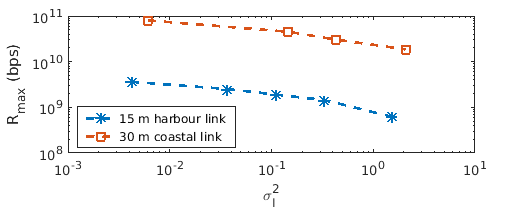}
	\caption{Maximum achievable data rate against $\sigma_I^2$ with a $20$~mW optical transmit power.}
	\label{capacity}
\end{figure}
Following the expressions described in Section~\ref{datameth}, the DGF fits can be used to find the maximum data rate supported by the channel. To find the maximum achievable rate, $R_{max}$ of a turbulent channel, we consider the worst case, i.e. $b_t=b_{tmax}$. This $R_{max}$ - calculated from~\eqref{Rmax}  with $L=10^5$ bits and $P_t=20$~mW for unipolar OOK - for 30~m coastal and 15~m harbour links are shown against $\sigma_I^2$ in Figure~\ref{capacity}. It shows that for both channels as $\sigma_I^2$ increases then the maximum achievable data rate decreases, this is expected due to both the lower total channel gain shown in the spatial distributions and the higher $D_{rms}^{DGF}$ shown in the temporal distributions for the $b_t$. As the total signal power arriving at the Rx decreases, it is also spread over a greater time compounding this drop in power by increasing ISI to further degrade performance. Despite the different $b_{tmax}$ required to yield a given $\sigma_I^2$, it is clear that the relative impact of $\sigma_I^2$ on $R_{max}$ is proportionally similar for both channels considered in this case study. The implication of this being that although $\sigma_I^2$ has a similar effect in different channel conditions, the higher $b_t$ required to cause those conditions means high turbulence can be considered unlikely in a more turbid channel.

Assuming an ideal Rx and transmitting using OOK modulation, then in coastal water over a 30~m link a maximum data rate of 80~Gbps can be attained when there is no turbulence. Similarly over a 15~m harbour link 3.8~Gbps is possible with no turbulence. When turbulence is included in the model then $R_{max}$ decreases in both channels and would have to be accounted for with higher link margin and/or channel equalisation. The maximum data rates achieved through these simulations show the potential of UOWC for high data rate transmission over short distances.
\section{Conclusion}\label{conc}
This paper presents a method of modelling turbulence in the UOWC channel as a scattering component, taking into account different water types rather than assuming generality for all water. Through this modelling technique the temporal and spatial impact of turbulence is highlighted. It is shown that as turbulence increases, the spatial and temporal spread of the channel also increase, but at different rates for different channel conditions. The distribution of received photon counts for the turbulent channel follows a Log-Normal distribution. The simulation method demonstrates that the impact of turbulence is greater when the channel has a lower number of scattering interactions per link. Furthermore it is shown that $\sigma_I^2$ is not general for all water conditions for a given parameter of turbulence induced scattering. This finding must be considered in future simulation works when modelling system performance in different, turbulent, water conditions. Finally the maximum data rate achievable in each water type is shown to decrease with an increase in $\sigma_I^2$. All these results combined provide evidence that if an UOWC channel is to be accurately modelled then turbulence cannot be examined in isolation from absorption and scattering.
\bibliographystyle{IEEEtran}
\bibliography{uwchannel_journal}
\end{document}